\documentclass[twocolumn,prl]{revtex4}
\usepackage{float}
\usepackage{amsfonts}
\usepackage{makeidx}
\usepackage{graphicx}
\usepackage{amsmath,amssymb,graphics,epsfig,color,times,bbm}

\begin{document}

\title{Multi-normal-mode splitting of a cavity in the presence of atoms -- towards the superstrong coupling regime}
\author{Xudong Yu$^{1}$, Dezhi Xiong$^{1}$, Haixia Chen$^{1}$, Pengjun Wang$^{1}$, Min Xiao$^{1,2}$, Jing Zhang$^{1}$$^{\dagger }$}
\affiliation{$^{1}$The State Key Laboratory of Quantum Optics and
Quantum Optics Devices, Institute of Opto-Electronics, Shanxi
University, Taiyuan 030006, P.R. China}

\affiliation{$^{2}$ Department of Physics, University of Arkansas,
Fayetteville, Arkansas 72701, USA}

\begin{abstract}
Multi-normal-mode splitting peaks are experimentally observed in a
system with Doppler-broadened two-level atoms inside a relatively
long optical cavity. In this system, the atoms-cavity interaction
can reach the ``superstrong coupling" condition with atoms-cavity
coupling strength $g\sqrt{N}$ to be near or larger than the cavity
free-spectral range $\Delta_{FSR}$. In such case, normal-mode
splitting can occur in many cavity longitudinal modes to generate
the multi-normal-mode splitting peaks, which can be well explained
by the linear dispersion enhancement due to the largely increased
atomic density in the cavity. Many new interesting phenomena might
come out of this superstrong atoms-cavity coupling regime.

\end{abstract}

\maketitle

Studies of strong coupling between atoms and cavity field have been
very active in the past 30 years \cite{Berman} due to its importance
in the fundamental understanding of quantum electrodynamics and
potential applications in quantum computation and quantum
information processing \cite{QC}. In the traditional cavity-quantum
electrodynamics (C-QED), high finesse microcavities are normally
used to enhance the single-photon coupling strength $g$
($=\sqrt{\frac{\mu^{2}\omega_{c}}{2\hbar V_{M}}}$, where
$\omega_{c}$ is the resonant frequency of the cavity, $\mu$ is the
atomic dipole matrix element, $V_{M}$ is the cavity mode volume), so
the strong-coupling condition of $g>\kappa,\gamma$ can be satisfied
even with a single atom \cite{Berman,one,two} (where $\kappa$ is the
cavity decay rate and $\gamma$ is the atomic decay rate). Two
normal-mode splitting peaks (i.e. Rabi sidebands) appear in the
cavity transmission spectrum due to such strong atom-cavity
interaction, with the frequency space between the two side peaks
given by $2g$. This atom-cavity coupling strength $g$ can be
enhanced by using an assemble of atoms to be $g\sqrt{N}$, where $N$
is the number of atoms in the cavity mode volume
\cite{eight1,eight2}. Normal-mode splitting in assemble of two-level
atoms has been demonstrated in atomic beams \cite{eight,nine}, cold
atomic cloud \cite{Hemmerich,ten,eleven} and Bose-Einstein
condensate \cite{twelve}. Recently, such normal-mode splitting was
even observed in Doppler-broadened two-level atoms in a hot atomic
vapor cell inside a low finesse ring cavity \cite{thirteen}.

A different regime of atoms-cavity interaction has been investigated
in a system with cold atoms in an optical lattice formed inside a
high-finesse, macroscopic optical cavity
\cite{Hemmerich,mech,scatter}. Interesting effects, such as
cavity-mediated collective light scattering due to the
self-organized atoms in the intracavity optical lattice and
collective atomic motion, were observed in such system
\cite{scatter}. Recently, normal-mode splitting and collective
mechanical effects have also been studied in such atoms-cavity
system \cite{Hemmerich}. The coherent backscattering between the two
propagating directions of a longitudinal mode has enhanced the
coupling between the atoms in the optical lattice and the cavity
fields, even when the fields are detuned far from the atomic
resonance \cite{scatter}.

In this Letter, we present our experimental demonstration of
multi-normal-mode splitting peaks due to strong coupling between
high-density, inhomogeneously-broadened two-level atoms and multiple
longitudinal cavity modes in a long Fabre-Perot optical cavity.
Here, we have reached a new regime of strong atoms-cavity
interaction, i.e. with $g\sqrt{N}>\Delta_{FSR}$, where
$\Delta_{FSR}$ is the free-spectral range (FSR) of the optical
cavity.  Meiser and Meystre theoretically studied a system under
such condition with a microscopic number of atoms \cite{thirteen1}
and named it as the "superstrong coupling" regime in the C-QED. Such
superstrong coupling condition has not been reached in previous
experiments due to either short optical cavity (therefore very large
$\Delta_{FSR}$) used to increase $g$ or relatively low atomic
density. In the current experimental system with
inhomogeneously-broadened two-level atoms in a long optical cavity,
the atomic density can be easily increased to satisfy this
superstrong coupling condition. With normal-mode splittings for
multiple longitudinal cavity modes, different cavity modes and their
atom-cavity polaritons can interact with each other, which can
provide the atom-mediated coupling between different cavity modes
via $\chi(3)$ nonlinearity \cite{kerr}. Such coupling between
different cavity field modes can be used for many applications in
quantum information processing, such as phase gates and multimode
entangled states. With such demonstrated new regime in atoms-cavity
interaction, many more interesting applications can be envisioned,
including ones in solid-state systems and photonic crystal cavities.

In the model of the coupled atoms-cavity system, the cavity is a
standing-wave cavity with length $L_{c}$, and the intracavity medium
is an assemble of two-level atoms, whose resonance frequency
$\omega_{a}$ is near a cavity resonance at frequency $\omega_{c}$.
The two-level atomic medium has a length $L_{a}<L_{c}$ with $N$
atoms in the cavity volume. The intensity transmission function of a
probe laser ($\omega_{L}$) for this coupled atoms-cavity system can
be easily found and is given by Eq.1 in Ref. \cite{eight}. The
frequency -dependent intensity-absorption coefficient and the
refractive index of the atomic medium are given by
\begin{eqnarray}
\alpha&=&\alpha_{0}\frac{\gamma_{a}^{2}}{4\Delta^{2}+\gamma_{a}^{2}}\label{absorb}
\\n&=&1-\alpha_{0}\frac{c}{\omega_{a}}\frac{2\Delta\gamma_{a}}{4\Delta^{2}+\gamma_{a}^{2}}, \label{refr}
\end{eqnarray}
respectively, where $\alpha_{0}=\omega_{a}N_{D}
|\mu|^{2}/\varepsilon_{0}\hbar c\gamma_{a}$ is the line-center
absorption coefficient. Notice that Eqs.1 and 2 are valid only when
$2\pi\alpha_{0}/\lambda_{a}\ll1$ \cite{thirteen2}. Here, the angle
sustained by the cavity mode is small, and the transverse decay rate
can be very closely approximated by the atomic free-space decay rate
$\gamma_{a}=\mu^{2}\omega_{a}^{3}/3\pi\hbar\epsilon_{0}c^{3}$. Thus,
the line-center absorption coefficient becomes
$\alpha_{0}=\frac{3\pi c^{2}}{\omega_{a}^{2}}N_{D}$.
$\Delta=\omega_{L}-\omega_{a}$ and
$\Delta_{ac}=\omega_{a}-\omega_{c}$ are the laser-atom and
atom-cavity frequency detunings, respectively. The cavity linewidth
is given by $\kappa=\Delta_{FSR}/F$, where $F$ is the cavity
finesse. This linear-dispersion theory was used to explain the
experimental observation in the system with a two-level atomic beam
passing through an one-centimeter long standing-wave cavity
\cite{eight}, where only one cavity longitudinal mode (m=0) was
considered.

Here, we can also apply this linear-dispersion theory to the case
with superstrong coupling, in which other cavity longitudinal modes
($m=\pm1$ and $m=\pm2$, et al) have to be included. We will only
consider the case of $\Delta_{ac}=0$. For an empty cavity, the
cavity transmission peaks are Lorenzian in shape and occur at
$\phi(\omega_{L})=\pm m2\pi$, where $\phi$ is the round-trip phase
shift experienced by the intracavity field going through the cavity
and $m=0,1,2,...$, with equal mode spaces given by $\Delta_{FSR}$,
as can be easily seen from the cavity transmission function (Eq. (1)
in Ref. \cite{eight}). With an intracavity (two-level) atomic
medium, the cavity transmission structure is significantly modified.
First, when the atomic density is high enough, so
$g\sqrt{N}\gg\gamma_{a}, \kappa$, but much smaller than
$\Delta_{FSR}$, $\phi(\omega_{L})=0$ will have two real solutions
due to the dispersion introduced by the atoms as given in Eq. (2).
This indicates that the center peak (m=0) in the cavity transmission
is split into two side peaks located at $\pm g\sqrt{N}$,
respectively, which is the standard normal-mode splitting, as shown
in Fig. 1(a). Under this condition, the other cavity modes ($m=\pm1$
and $m=\pm2$, et al) are not affected by the atoms and still have
equal spaces between them given by $\Delta_{FSR}$. However, as the
condition that $g\sqrt{N}$ is near or larger than $\Delta_{FSR}$ is
satisfied, then not only the center cavity mode ($m=0$) has
normal-mode splitting, other cavity modes (such as $m=\pm1$) will
interact with the atoms and have their own normal-mode splitting
peaks (i.e. $\phi(\omega_{L})=2\pi$ or $\phi(\omega_{L})=-2\pi$ will
also have two real solutions), as shown in Fig.1(b). The positions
of their two normal-mode splitting peaks for $|m|\geq1$ locate in
the two sides of the atomic resonant frequency and present
asymmetric structure as shown in Fig. 1(b). To see this, let's
consider only the $m=1$ cavity mode, which can be viewed as a mode
with an effective cavity detuning of $\Delta_{FSR}$. As it interacts
with atoms under the ``superstrong coupling" condition, the Rabi
sidebands become very asymmetric (due to large effective detuning),
with one large peak on the right (near the original empty cavity
peak position) of the atomic resonance and a smaller peak on the
left of the atomic resonance (all labeled as ``1" in Figs.
1(b)-1(d)). When the atomic density gets even higher, more cavity
modes (such as $m=\pm2$ and $m=\pm3$, et al) will participate in the
mode-splitting process, which form the multi-normal-mode splitting
structure, as shown in Figs. 1(c) and 1(d).

For our experimental situation, we consider a system with two-level
rubidium atoms (in a vapor cell) inside an optical standing-wave
cavity of 17.7 cm long. Here, $\gamma_{a}=2\pi\times6 MHz$. The
original normal-mode splitting $g\sqrt{N}$ can reach and even be
larger than $\Delta_{FSR}$ by increasing the temperature of the
atomic cell (the corresponding atomic density is increased quickly
with temperature). Due to Doppler effect, the absorption (and
therefore also the dispersion) profile is much broader with a width
of $\delta \omega_{D}=\frac{\omega_{c}}{c}\sqrt{\frac{2k_{B}T}{m}}$.
For simplicity in discussion, we can replace the homogeneous
absorption linewidth $\gamma_{a}$ by the Doppler width $\delta
\omega_{D}$ in Eqs. 1 and 2, which is approximately valid. At the
same time, the line-center absorption coefficient is changed into
$a_{0}=\frac{3\pi
c^{2}}{\omega_{a}^{2}}\frac{\gamma_{a}}{\omega_{D}}N_{D}$. In order
to accurately calculate the absorption and dispersion properties in
the Doppler-broadened atomic system, we need to replace Eqs.1 and 2
by Eqs.5 and 6 of Ref.\cite{thirteen}. Notice that the second-term
in the index of refraction of the intracavity medium (Eq. 2) depends
on the atomic density in the cavity ($a_{0}\propto N_{D}$). So, as
the temperature of the atomic cell increases, more cavity modes will
participate in the mode-splitting interactions to generate the
multi-normal-mode splitting peaks.

Parameters used in plotting Fig. 1 are basically the same as in our
experiment. For very low absorption coefficient $a_{0}$
(corresponding to low temperature or atomic density), the Doppler
absorption and dispersion width ($\delta \omega_{D}=2\pi\times343
MHz$) is smaller than the cavity free-spectral range
($\Delta_{FSR}=2\pi\times850 MHz$). Therefore, except for the two
middle normal-mode peaks, all other peaks are the cavity
transmission peaks without mode-spitting (corresponding to Fig.
1(a)), since they do not interact strongly with the atoms. The
shorter transmission peak height of the two middle normal-mode peaks
is due to the atomic absorption. As the absorption coefficient
$a_{0}$ gets larger (higher temperature), the dispersion changes and
more FSR cavity modes join the mode-splitting interactions, as shown
in Figs. 1(b)-1(d). Figures 1(e) and 1(f) (which are the re-plots of
Figs.1(b) and (d), respectively, with the FSR cavity mode number $m$
as the horizontal axis) give a more clear insight into the positions
and heights of the multi-normal-mode splitting peaks. Such plots are
the typical avoided-crossing plots commonly used in cavity-QED.

%
\begin{figure}
\centerline{
\includegraphics[width=3.8in]{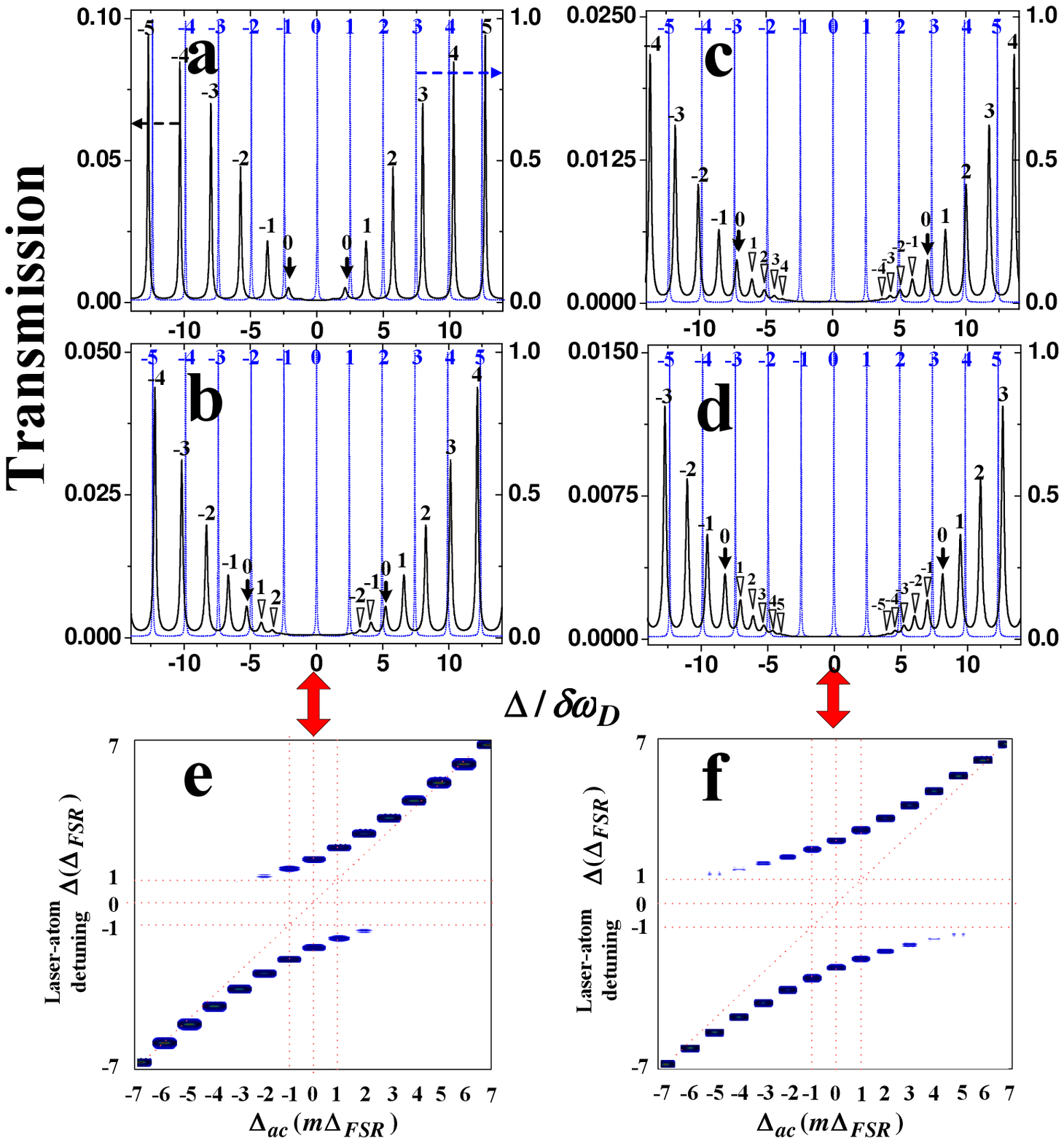}
} \vspace{0.1in}
\caption{(Color online). Theoretical calculations of the
transmission spectra of the coupled atoms-cavity system with the
Doppler-broadened two-level atoms with different absorption
coefficients. For comparison, the cavity transmission spectrum for
the empty cavity (blue dashed) is plotted in (a), (b), (c) and (d).
The vertical scale is normalized to the light intensity transmitted
through the empty cavity. This is for an optical cavity with
$\Delta_{FSR}=2\pi\times 850$ MHz. $m=0,\pm1,\pm2,...$ label the
multi-normal-mode splitting peaks coming from the FSR cavity modes.
The finesse of the empty cavity $F=20$ and $\lambda=780$ nm. (a)
$a_{0}L_{a}=12$ (corresponding to
$N_{D}L_{a}=9.4\times10^{15}/m^{2}$); (b) $a_{0}L_{a}=70$
($N_{D}L_{a}=5.5\times10^{16}/m^{2}$); (c) $a_{0}L_{a}=130$
($N_{D}L_{a}=1.0\times10^{17}/m^{2}$); (d) $a_{0}L_{a}=170$
($N_{D}L_{a}=1.3\times10^{17}/m^{2}$); (e) and (f) are the re-plots
of (b) and (d), respectively. \label{Fig1} }
\end{figure}

The experiment was done using a standing-wave cavity of 17.7 cm
long, as shown in Fig.2. The cavity composes of two mirrors with
same radius of curvature of 100 mm. The reflectivity is $90\%$ at
780 nm for the input coupler $M1$, which is mounted on a PZT to
adjust the cavity length. The output coupler $M2$ has a reflectivity
of $99.5\%$ at 780 nm. The finesse of the cavity including the
losses of two faces of the atomic cell is about $F=20$. The length
of the vapor cell is 5 cm. The temperature of the vapor cell can be
controlled by a heater. A grating-stabilized diode laser, which
first passes a standard polarization maintaining single-mode fiber,
was used as the cavity input beam with an input power of 4 mW. The
laser beam with the spatial mode filter by an optical fiber is easer
to mode-match to the $TEM_{00}$ mode of the optical cavity. The
diode laser has also been divided into several parts to be used in
the saturated absorption spectroscopy (using another rubidium atomic
cell) and in the F-P cavity for monitoring the frequency and mode of
the laser. The optical cavity length was kept unchanged and the
input laser frequency was scanned to measure the transmission
spectra. The transition between the $5S_{1/2}$, $F=1$ and
$5P_{3/2}$, $F'$ in $^{87}Rb$ was used for the two-level system.

%
\begin{figure}
\centerline{
\includegraphics[width=3in]{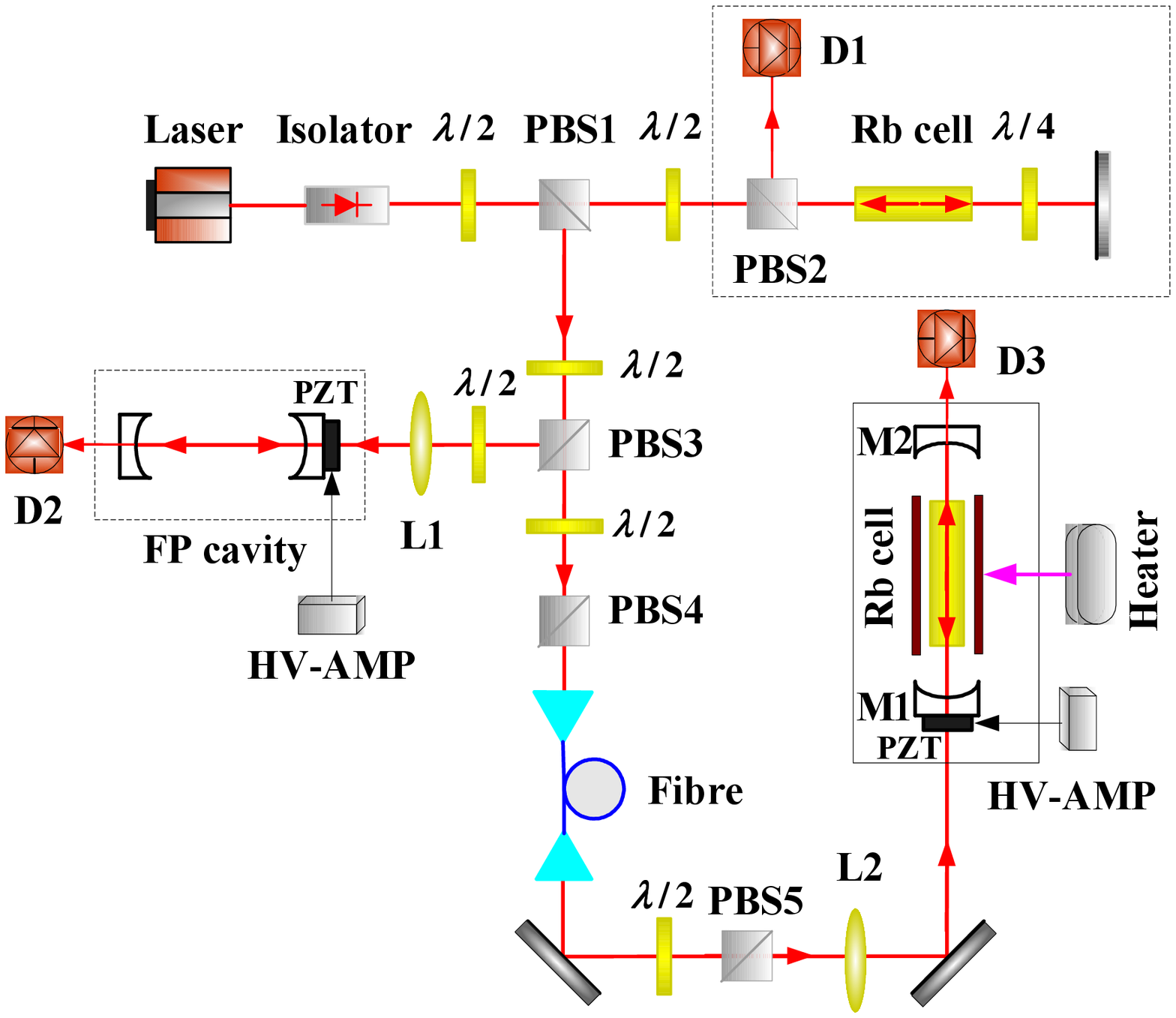}
} \vspace{0.1in}
\caption{(Color online). Schematic of the experimental setup of the
coupled atoms-cavity system. $\lambda /2$: half-wave plate;
$D1,D2,D3$: detectors; HV-AMP: high voltage amplifier; PZT:
piezoelectric transducer; PBS: polarized beam splitter; $L1, L2$:
optical lens. \label{Fig2} }
\end{figure}

At a low atomic cell temperature ($T=105^{0}C$, corresponding to
$N_{D}L_{a}=2.5\times10^{17}/m^{2}$ \cite{steck}), the cavity
transmission spectrum is given in Fig.3(a). At both sides of the
resonant frequency of atoms, the atoms-cavity normal modes can be
observed. Only two normal-mode splitting peaks appear in the
transmission spectrum, and the other outside peaks are simply the
cavity FSR peaks without mode-splitting, as predicted in Fig.1(a).
Notice that the atomic density predicted theoretically in Fig.1(a)
is about an order of magnitude smaller than the density predicted by
Ref. \cite{steck} for rubidium at $T=105^{0}C$, which is mainly due
to a large fraction of atoms populated at $5S_{1/2}$, $F=2$ when
without a pump light \cite{forteen}. As the temperature increases
($T=110^{0}C$), more FSR cavity modes participate in the
mode-splitting process and generate the multi-normal-mode splitting
peaks, as shown in Fig.3(b), which also agrees with the theoretical
curve of Fig.1(b). In order to label the two original normal-mode
splitting peaks, we must monitor the peaks carefully by slowly
increasing the temperature. The asymmetry in the two transmission
side peaks compared with Fig.1 is due to the influence of absorption
from $5S_{1/2}$, $F=2$ to $5P_{3/2}$, $F'$ transition in $^{85}Rb$,
as shown in Fig.3(e). As the temperature is further increased
($T=115^{0}C$ and $T=120^{0}C$), more FSR cavity modes join in the
mode-splitting interactions as shown in Figs.3 (c) and (d) (which
are also predicted in Figs.1(c) and 1(d)) due to the largely
increased dispersion as indicated by Eq.2. Since the atomic density
(and therefore the absorption coefficient $a_{0}$) depends
exponentially on the temperature, the dispersion change (and
therefore the number of FSR cavity modes generating
multi-normal-mode splitting) has a very sensitive dependence on the
temperature of the atomic cell, as shown in Fig.3.

%
\begin{figure}
\centerline{
\includegraphics[width=3.8in]{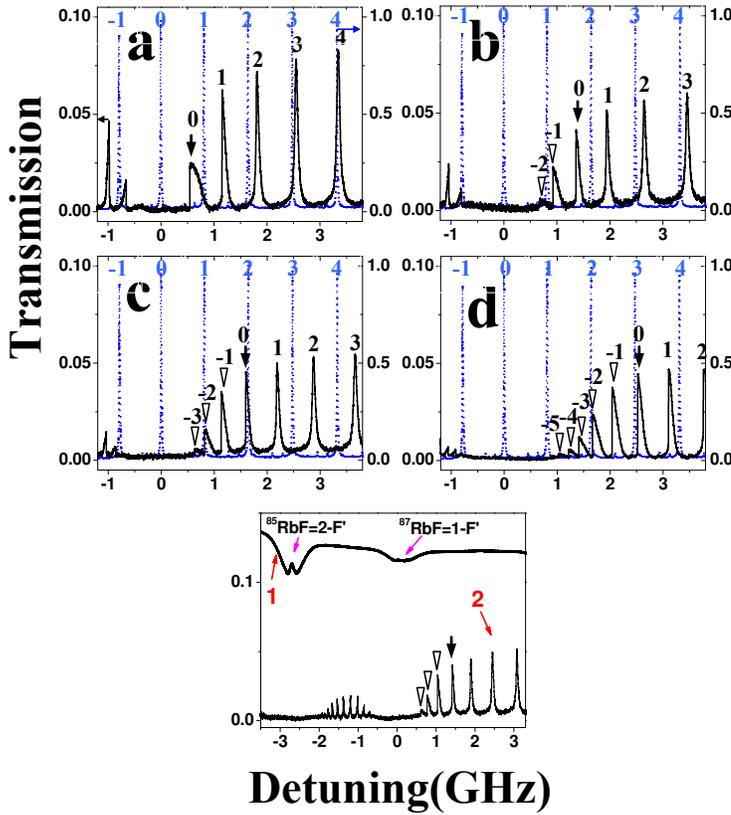}
} \vspace{0.1in}
\caption{(Color online). Experimentally measured transmission
spectra of the coupled atoms-cavity system with the
Doppler-broadened two-level atoms at different temperatures of the
atomic cell. The curves (a), (b), (c) and (d) correspond to (e) with
a narrow frequency detuning range. For comparison, the cavity
transmission spectrum for the empty cavity (blue dashed) is added in
(a), (b), (c) and (d). The vertical scale is normalized to the light
intensity transmitted through the empty cavity. (a) $T=105^{0}C$;
(b) $T=110^{0}C$; (c) $T=115^{0}C$; (d) $T=120^{0}C$; (e) The
saturated absorption spectroscopy (curve 1) and the corresponding
cavity transmission spectra (curve 2) of the coupled atoms-cavity
system. $T=120^{0}C$. \label{Fig3} }
\end{figure}

In summary, we have studied the cavity transmission spectra in a
system with Doppler-broadened two-level atoms in a standingwave
cavity under the ``superstrong coupling" condition of $g\sqrt{N}$
larger or equal to $\Delta_{FSR}$. In this superstrong atoms-cavity
coupling region, mode-splitting occurs in many FSR cavity modes due
to the interactions with the intracavity Doppler-broadened atoms,
and multi-normal-mode splitting peaks were observed experimentally.
Such multi-normal-mode peaks depend sensitively on the atomic
density. This phenomenon can be qualitatively explained by using the
linear absorption and dispersion theory of the cavity transmission,
and by taking into account the sensitive dependence of the index of
refraction of the intracavity medium on the atomic density. This
work sheds new light on the investigations of this novel
``superstrong coupling" region in the exciting field of collective
coupling between atoms and cavity, and can lead to interesting
applications in quantum information processing.

$^{\dagger} $Corresponding author's email address:
jzhang74@sxu.edu.cn, jzhang74@yahoo.com


J. Zhang thanks K. Peng, C. Xie and T. Zhang for the helpful
discussions. This research was supported in part by NSFC for
Distinguished Young Scholars (Grant No. 10725416), National Basic
Research Program of China (Grant No. 2006CB921101), NSFC Project for
Excellent Research Team (Grant No. 60821004), and NSFC (Grant No.
60678029).



\begin{thebibliography}{99}
\bibitem{Berman} P. R. Berman, Cavity Quantum Electrodynamics
(Advances in Atomic, Molecular, and Optical Physics, Academic, New
York, 1994).

\bibitem{QC} T. Pellizzari, S. A. Gardiner, J. I. Cirac, P. Zoller,
Phys. Rev. Lett. \textbf{75}, 3788 (1995); L. M. Duan, H. J. Kimble,
Phys. Rev. Lett. \textbf{92}, 127902 (2004); J. I. Cirac, P. Zoller,
H. J. Kimble, H. Mabuchi, Phys. Rev. Lett. \textbf{78}, 3221 (1997).

\bibitem{one} A. Boca, R. Miller, K. M. Birnbaum, A. D. Boozer, J. McKeever, and H. J. Kimble, Phys. Rev. Lett. \textbf{93}, 233603 (2004).

\bibitem{two} P. Maunz, T. Puppe, I. Schuster, N. Syassen, P. W. H. Pinkse, and G. Rempe, Phys. Rev. Lett. \textbf{94}, 033002 (2005);
 T. Puppe, I. Schuster, A. Grothe, A. Kubanek, K. Murr, P. W. H. Pinkse, and G. Rempe, Phys. Rev. Lett. \textbf{99}, 013002 (2007).

\bibitem{eight1} M. Tavis, F. W. Cummings, Phys. Rev. \textbf{170}, 379
(1968).

\bibitem{eight2} G. S. Agarwal, Phys. Rev. Lett. \textbf{53}, 1732 (1984).

\bibitem{eight} Y. Zhu, D. J. Gauthier, S. E. Morin, Q. Wu, H. J. Carmichael, and T. W. Mossberg, Phys. Rev. Lett. \textbf{64}, 2499 (1990).

\bibitem{nine} R. J. Thompson, G. Rempe, and H. J. Kimble, Phys. Rev. Lett. \textbf{68}, 1132 (1992).

\bibitem{Hemmerich}J. Klinner, M. Lindholdt, B. Nagorny, A.
Hemmerich, Phys. Rev. Lett. \textbf{96}, 023002 (2006).

\bibitem{ten} A. K. Tuchman, R. Long, G. Vrijsen, J. Boudet, J. Lee, and M. A. Kasevich, Phys. Rev. A \textbf{74}, 053821 (2006).

\bibitem{eleven} G. Hernandez, J. Zhang, and Y. Zhu, Phys. Rev. A \textbf{76}, 053814 (2007).

\bibitem{twelve} S. Gupta, K. L. Moore, K. W. Murch, and D. M. Stamper-Kurn, Phys. Rev. Lett. \textbf{99}, 213601 (2007);
Y. Colombe, T. Steinmetz, G. Dubois, F. Linke, D. Hunger, J.
Reichel, Nature \textbf{450}, 272 (2007); F. Brennecke, T. Donner,
S. Ritter, T. Bourdel, M. Kohl, T. Esslinger, Nature \textbf{450},
268 (2007).

\bibitem{thirteen} J. Gea-Banacloche, H. Wu, and M. Xiao, Phys. Rev.
A \textbf{78}, 023828 (2008).

\bibitem{mech} P. Domokos and H. Ritsch, Phys. Rev. Lett. \textbf{89}, 253003
(2002); J. K. Asboth, P. Domokos, H. Ritsch, and A. Vukics, Phys.
Rev. A \textbf{72}, 053417 (2005).

\bibitem{scatter}B. Nagorny, Th. Elsasser, A. Hemmerich, Phys. Rev. Lett. \textbf{91}, 153003
(2003); D. Kruse, C. von Cube, C. Zimmermann, and P. W. Courteille,
Phys. Rev. Lett. \textbf{91}, 183601 (2003); A. T. Black, H. W.
Chan, and V. Vuletic, Phys. Rev. Lett. \textbf{91}, 203001 (2003);
S. Slama, S. Bux, G. Krenz, C. Zimmermann, and Ph. W. Courteille,
Phys. Rev. Lett. \textbf{98}, 053603 (2007).


\bibitem{thirteen1} D. Meiser and P. Meystre, Phys. Rev. A \textbf{74}, 065801 (2006).

\bibitem{kerr} L. M. Duan, J. I. Cirac, P. Zoller, E. S. Polzik,
Phys. Rev. Lett. \textbf{85}, 5643 (2000); L. M. Duan, G. Giedke, J.
I. Cirac, P. Zoller, Phys. Rev. Lett. \textbf{84}, 4002 (2000); B.
Julsgaard, A. Kozhekin, E. S. Polzik, Nature \textbf{413}, 400
(2001).

\bibitem{thirteen2} R. W. Boyd, Nonlinear Optics, (Academic, San Diego, CA, 2003).

\bibitem{steck} http://steck.us/alkalidata.

\bibitem{forteen} H. Wu, J. Gea-Banacloche, and M. Xiao,
Phys. Rev. Lett. \textbf{100}, 173602 (2008).


\end{thebibliography}
\end{document}